\documentclass[10pt,conference]{IEEEtran}
\IEEEoverridecommandlockouts
\usepackage{etoolbox}
\usepackage[most]{tcolorbox}
\newtcolorbox{mybox}[1][]{
breakable,
  arc=1mm,
  boxrule=1pt,
  colback=yellow!14,
  colframe=black!80,
  fonttitle=\bfseries,
  title=#1,
  left=1mm,
  right=1mm,
  top=1mm,
  bottom=1mm
}
\usepackage{cite}
\usepackage{subfigure}
\usepackage{amsmath,amssymb,amsfonts}
\usepackage{algorithmic}
\usepackage{graphicx}
\usepackage{textcomp}
\usepackage{xcolor}
\usepackage{cases}
\usepackage{multirow}
\usepackage{booktabs}
\usepackage[capitalize]{cleveref}
\usepackage{makecell}
\usepackage{color, soul}
\usepackage{bbding}
\usepackage{colortbl}
\usepackage{arydshln}
\usepackage{url}
\usepackage[group-separator={,}]{siunitx}
\usepackage[most]{tcolorbox}
\def\BibTeX{{\rm B\kern-.05em{\sc i\kern-.025em b}\kern-.08em
    T\kern-.1667em\lower.7ex\hbox{E}\kern-.125emX}}
\DeclareMathOperator*{\argmax}{arg\,max}
\begin{document}

\title{LogLM: From Task-based to Instruction-based Automated Log Analysis}

\author{\IEEEauthorblockN{Yilun Liu$^{1}$, Yuhe Ji$^{1}$, Shimin Tao$^{1}$, Minggui He$^{1}$, Weibin Meng$^{1}$, Shenglin Zhang$^{2}$ \\ Yongqian Sun$^{2}$, Yuming Xie$^{1}$, Boxing Chen$^{3}$, Hao Yang$^{1}$}
\IEEEauthorblockA{$^{1}$Huawei, China}
\IEEEauthorblockA{$^{2}$Nankai University, China}
\IEEEauthorblockA{$^{3}$Huawei Canada, Canada}
\IEEEauthorblockA{\{liuyilun3, jiyuhe1, taoshimin, heminggui, mengweibin3, yuming.xie, boxing.chen,}
\IEEEauthorblockA{yanghao30\}@huawei.com, \{zhangsl, sunyongqian\}@nankai.edu.cn}}

\maketitle

\begin{abstract}
Automatic log analysis is essential for the efficient Operation and Maintenance (O\&M) of software systems, providing critical insights into system behaviors. However, existing approaches mostly treat log analysis as training a model to perform an isolated task ( \emph{e.g.}, anomaly detection, log parsing, \emph{etc.}) using task-specific log-label pairs. These task-based approaches are inflexible in generalizing to complex scenarios, depend on task-specific training data, and cost significantly when deploying multiple models. In this paper, we propose an instruction-based training approach that transforms log-label pairs from multiple tasks and domains into a unified format of instruction-response pairs. Our trained model, LogLM, can follow complex user instructions and generalize better across different tasks, thereby increasing flexibility and reducing the dependence on task-specific training data. By integrating major log analysis tasks into a single model, our approach also relieves model deployment burden. Experimentally, LogLM outperforms existing approaches across five log analysis capabilities, and exhibits strong generalization abilities on complex instructions and unseen tasks.
\end{abstract}


\section{Introduction}\label{sec:intro}
In the ever-evolving landscape of software systems, log analysis has become a critical component of system Operation and Maintenance (O\&M). Logs serve as an invaluable source of information, capturing events, errors, and other significant activities that can shed light on system performance and potential failures. However, as systems grow in complexity, manually analyzing logs becomes inefficient and requires advanced expertise. Thus, automated log analysis has emerged as a solution, using algorithms and machine learning models to extract meaningful insights from log data. This automation enhances efficiency, enabling timely detection of anomalies, identifying potential patterns, and aiding O\&M engineers in taking actions, thereby helping maintain system reliability~\cite{oliner2012advances}.

Traditionally, the research field of log analysis has been divided into discrete tasks, each with its own methodologies and benchmarking datasets. Two of the most widely studied tasks are log parsing and log-based anomaly detection. Log parsing involves transforming raw, unstructured log data into a structured format, with event templates and key variables extracted from log messages~\cite{zhu2019tools}. Anomaly detection, on the other hand, focuses on identifying deviations from normal log behavior, which may signal underlying issues within the system~\cite{le2022log}. Beyond these, other studied tasks include log interpretation~\cite{liu2024logprompt}, which aims to make logs more understandable to human, root cause analysis~\cite{chen2024automatic}, which seeks to reveal the causes of system failures from logs, and solution recommendation~\cite{ahmed2023recommending}, which suggests corrective actions for the identified problems in logs.

Recently, with the rapid advancements in AI technologies, particularly in large language model (LLM), a variety of LLM-based approaches for log analysis have been introduced~\cite{jiang2024lilac,pan2024raglog,zhong2024logparser}. LLMs such as GPT-4~\cite{achiam2023gpt} and Claude-3.5~\cite{claude} are capable of following complex instructions and performing a wide range of tasks. However, as shown in Fig.~\ref{fig_hook}(a), most of these approaches continue to treat log analysis as isolated tasks, training LLMs to handle standalone tasks such as log parsing or anomaly detection without considering the interrelations between these tasks. We observed that this task-based training approach fails to fully capitalize on the capabilities of LLMs (\emph{e.g.}, a performance drop of 8.70\% in log parsing and 69.47\% in anomaly detection in Fig.~\ref{ablation_exp}). For LLMs nowadays, the training paradigm is instruction-based (called instruction tuning~\cite{zhang2023instruction}), involving training on natural instructions across multiple tasks and domains, thereby offering greater flexibility and broader application. Furthermore, current task-based methods encounter several limitations when applied to real-world industrial scenarios:

\begin{figure}[!t]
 \centering  
 \subfigbottomskip=-2pt 
 \subfigcapskip=-2pt 
 \subfigure[Task-based log analysis]{
  \includegraphics[width=0.96\linewidth]{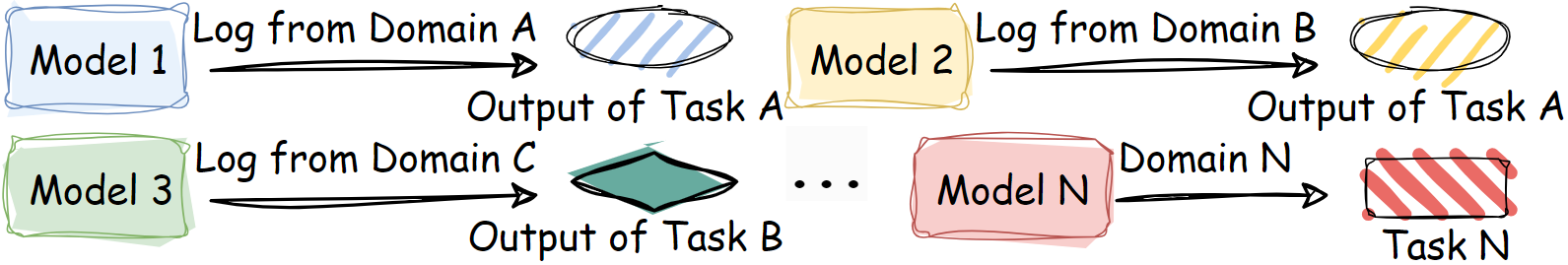}}
   \\
 \subfigure[Instruction-based log analysis]{
  \includegraphics[width=0.96\linewidth]{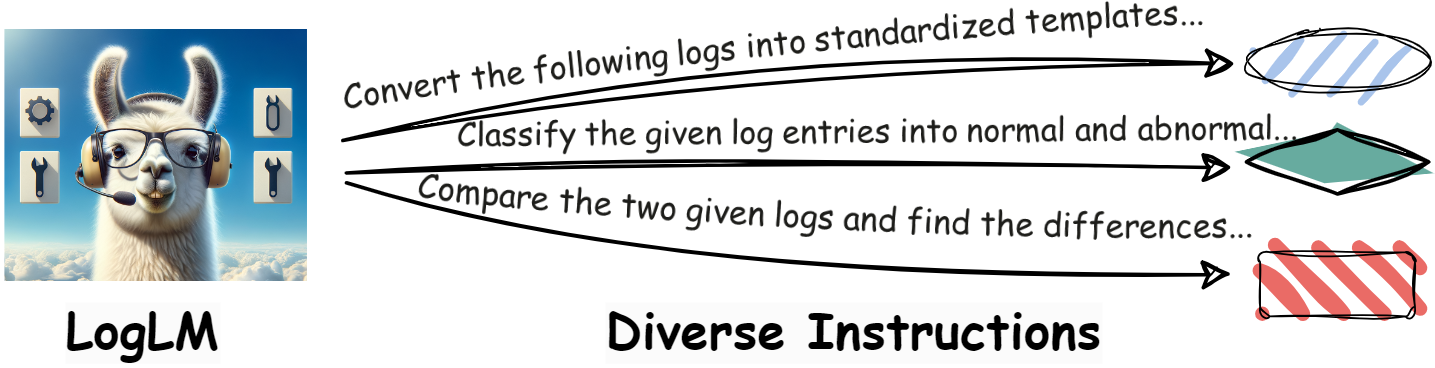}}
 \caption{Illustrated comparison between (a) existing task-based log analysis approaches and (b) our LogLM, an instruction-based log analysis model.}
\label{fig_hook}
\end{figure}

\textbf{(1) Inflexibility in generalizing to unseen tasks and complex user instructions.} Task-based log analysis models are typically trained to output predictions on a fixed input format, which leads to an inflexible human-machine interactivity style in responding to real-world user instructions, particularly for dynamic O\&M environments where new types of tasks often emerge. For example, in real-world cases, instead of parsing a log into a template, an O\&M engineer may want the model to compare two logs and find the differences~\cite{How_to_Use_ChatGPT_to_Compare_Documents}. In addition to reporting anomaly in logs, the engineer may instruct the model to further output a justification along with its conclusion~\cite{liu2024logprompt}. Existing methods are designed to handle a fixed task (\emph{e.g.}, input a log and output ``abnormal'' or ``normal'') and struggle to adapt to such flexible instructions or unfamiliar tasks, leading to a limited user experience.

\textbf{(2) Dependency on task-specific and domain-specific training data caused by ignoring cross-task (or domain) connections.} As observed by recent studies~\cite{liu2024interpretable,tao2023biglog,tao2022logstamp}, existing approaches are dependent on task-specific training data. Their performances can drop significantly in online situation, where available task-specific logs in the specific domain (\emph{e.g.}, modules, versions and devices) is scarce due to frequent software update and third-party module import~\cite{liu2024interpretable}. The dependency on task-specific or domain-specific training data is due to the ignorance of inherent connections between different analysis tasks and log domains. For example, accurately parsing a log’s structure can significantly improve anomaly detection~\cite{fu2023empirical}; a correct diagnosis of the root cause of a failure is crucial for generating a meaningful solution~\cite{ahmed2023recommending}. Also, logs from different domains are heterogeneous in format and structure, but may share some common patterns~\cite{tao2023biglog}. Just as human experts draw from prior knowledge across different domains, LLMs may also apply insights from solving one log analysis task to another and enriches analysis abilities by integrating log patterns from diverse domains, thereby requiring less in-domain and task-specific training data.

\textbf{(3) Significant burden in deployment.} In large-scale software systems, the sheer volume of O\&M tasks and diversity of logs from different domains demand that engineers deploy multiple specialized models, each fine-tuned for a particular task or domain. This leads to increased complexity in managing these models, as hundreds of specialized models may need to be maintained (\emph{e.g.}, five tasks and nine domains leads to 45 fine-tuned models to deploy), which is also noted as the ``pipeline challenge''~\cite{The_State_of_LLM_Operations_or_LLMOps}. Such an approach not only consumes significant computational resources but also burdens engineers with managing and updating a vast array of models.

\begin{figure}[t!]
    \centering
  \includegraphics[width=\linewidth]{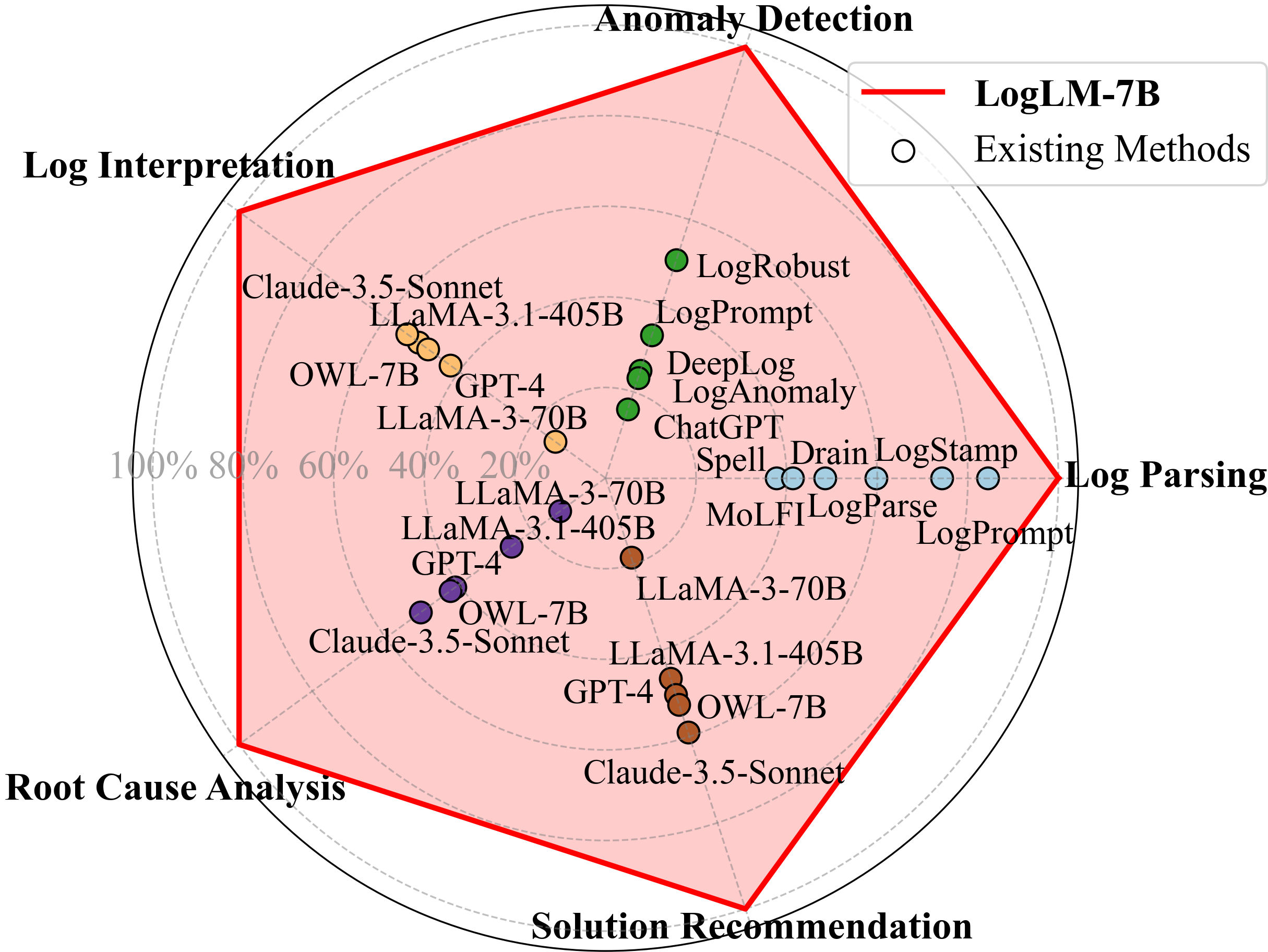}
  \caption{Comparison of average performance between LogLM and existing methods across five log analysis capabilities. The dots indicate the relative percentage of baselines' average performances in comparison to LogLM's. LogLM-7B is fine-tuned from an open-source LLM with 7B parameters.}
  \label{fig_radar}
\end{figure}
In this paper, we propose LogLM, a novel instruction-based approach to log analysis that addresses these limitations. (Strictly, this is moving from ``task-based'' models to an ``instruction-tuned unified-task'' model.) As shown in Fig.~\ref{fig_hook}(b), our approach transforms task-specific log-label pairs into instruction-response pairs, allowing the model to learn to properly respond to diverse user instructions. Instead of focusing on performing isolated tasks, our goal is to build a model that has comprehensive log analysis abilities. Thus, LogLM is trained on diverse instruction pairs, covering five log analysis capabilities and spanning nine log domains. Compared with existing methods, LogLM offers several key advantages: \textbf{(1) Instruction-following capability enables flexibility.} By training on natural instructions, LogLM can handle complex user instructions (Fig.~\ref{fig_user_instruction}), providing greater flexibility. \textbf{(2) Deep understanding of cross-task (or domain) connections relieves dependency.} Our training set breaks the barriers among different analysis tasks and log domains, enabling LogLM possessing deeper understanding of connections between tasks and domains. When in-domain (or task) training data is scarce, LogLM learns from out-of-domain (or task) data, as indicated by the increased performance in Fig.~\ref{ablation_exp} after introducing instructions from other tasks and domains in the training set. \textbf{(3) Single-model deployment reduces burden.} LogLM's multi-task ability allows for deploying only a single model for complex systems. Our contributions are:
\begin{itemize}
    \item We present a new approach to log analysis that uses instruction tuning to combine cross-task and cross-domain log data into a standard instruction-response format, allowing for more effective log analysis capabilities. In Fig.~\ref{fig_radar}, our model aces across all five analysis abilities.
    \item We develop a highly generalized instruction-following model capable of handling complex user instructions and performing previously unseen tasks (achieving best in 10 out 12 test terms on unseen tasks as shown in Table~\ref{tab:unseen_task}).
    \item We open-source the full training dataset and codes of LogLM, facilitating the development of future models.\footnote{Dataset and code available at \url{https://github.com/lunyiliu/LogLM}}
\end{itemize}

\section{Related Work}
\subsection{LLMs \& Instruction Tuning}

Recent LLMs, such as GPT-4~\cite{achiam2023gpt}, have demonstrated the capability to execute complex tasks and generate appropriate responses to human instructions~\cite{kalyan2023survey,liu2024coachlm,kung2023performance}. The development of these capabilities follows a two-stage process. The first stage, known as pre-training, involves training a foundation model to predict subsequent words in large-scale corpora \cite{brown2020language}. Despite foundation models like LLaMA \cite{touvron2023llama} being capable of sentence completion, they exhibit limitations in effectively responding to human instructions. To address this deficiency, LLMs undergo a second phase termed human alignment. A common method in this phase is instruction tuning~\cite{wei2021finetuned}, where the foundation model is fine-tuned using a diverse set of human instructions involving various tasks, paired with corresponding desired responses. This process aims to align the model’s outputs with human expectations and enhance its ability to generalize to unseen instructions~\cite{raffel2020exploring,wei2021finetuned,mishra2022cross}. It leverages instruction datasets composed of structured instruction pairs. Each pair consists of an \textsc{Instruction}, representing a human input, and a \textsc{Response}, representing the ideal output that resolves the instruction's task. Through training on these pairs, the model learns to apply its pre-trained knowledge to generate appropriate responses to varied instructions and can generalize to new instructions~\cite{zhang2023instruction}.

Our work, LogLM, can be seen as a pioneering exploration on leveraging the technique of instruction tuning to solve industrial challenges in the field of log analysis. We also noticed a similar work using LLMs to analyze other semi-structured data similar to log data~\cite{Table-GPT}, where diverse instructions are synthesized for multiple table-based tasks.
\subsection{Log Analysis}
\subsubsection{Task of Log Parsing}

Log parsing is a widely studied task designed to reduce the volume of extensive log data by extracting meaningful templates from raw logs, facilitating further analysis tasks such as anomaly detection. Early approaches to log parsing primarily focused on coarse-level parsing, which involves aggregating lexically similar logs into templates. A log template only retains static parts in a log and replaces dynamic variables with a symbol of \verb|<*>|. Examples of these methods include cluster-based~\cite{zhu2019tools,fu2009execution}, heuristic~\cite{du2016spell,makanju2009clustering}, and tree-based methods~\cite{zhang2017syslog,he2017drain,chu2021prefix}.

Beyond coarse-level methods, there is a growing body of research dedicated to fine-level log parsing, which emphasizes the significance of variables within logs~\cite{meng2020logparse,tao2022logstamp,huo2023semparser,li2023did}. 

Additionally, recent methods have begun to directly leverage LLMs for log parsing, such as LogPPT~\cite{le2023log} and LogPrompt~\cite{liu2024interpretable}, while other methods utilize LLMs as an enhancement to existing models~\cite{jiang2024lilac,zhong2024logparser}.

\subsubsection{Task of Log-based Anomaly Detection}\label{sec:anomaly_relatedwork}

Anomaly detection involves distinguishing anomalies within input logs. Following Le~\emph{et al.}~\cite{le2022log}, anomaly is defined as a pattern that violates normal sessions of systems. Techniques for anomaly detection can be categorized into two main classes: session-level methods and template-level methods. Both approaches typically begin by parsing raw logs into templates to facilitate analysis. Session-level methods subsequently aggregate log templates into sessions using a windowing strategy, and treat these log sessions as the minimum unit for anomaly detection. In contrast, template-level methods directly identify anomalies based on the semantics of individual log templates. If any template within a log session is predicted to be abnormal, template-level methods classify the entire session as anomaly.

Within session-level methods, further distinctions can be made between forecast-based and classification-based approaches. Classification-based methods need both normal and abnormal training logs for training~\cite{zhang2019robust,lu2018detecting,le2021log}, whereas forecast-based methods require only normal historical logs~\cite{du2017deeplog,meng2019loganomaly}. 

Template-level methods include techniques such as LogPrompt~\cite{liu2024interpretable}, which designs chain-of-thought prompts to detect anomalies in log entries using ChatGPT, and RAGLog~\cite{pan2024raglog}, which employs retrieval-augmented generation to assist anomaly detection using LLMs. Additionally, Cui~\emph{et al.}~\cite{cui2024logeval} proposed a benchmark for template-level anomaly detection and evaluates various LLMs on this task.

\begin{figure*}[t!]
    \centering
  \includegraphics[width=\linewidth]{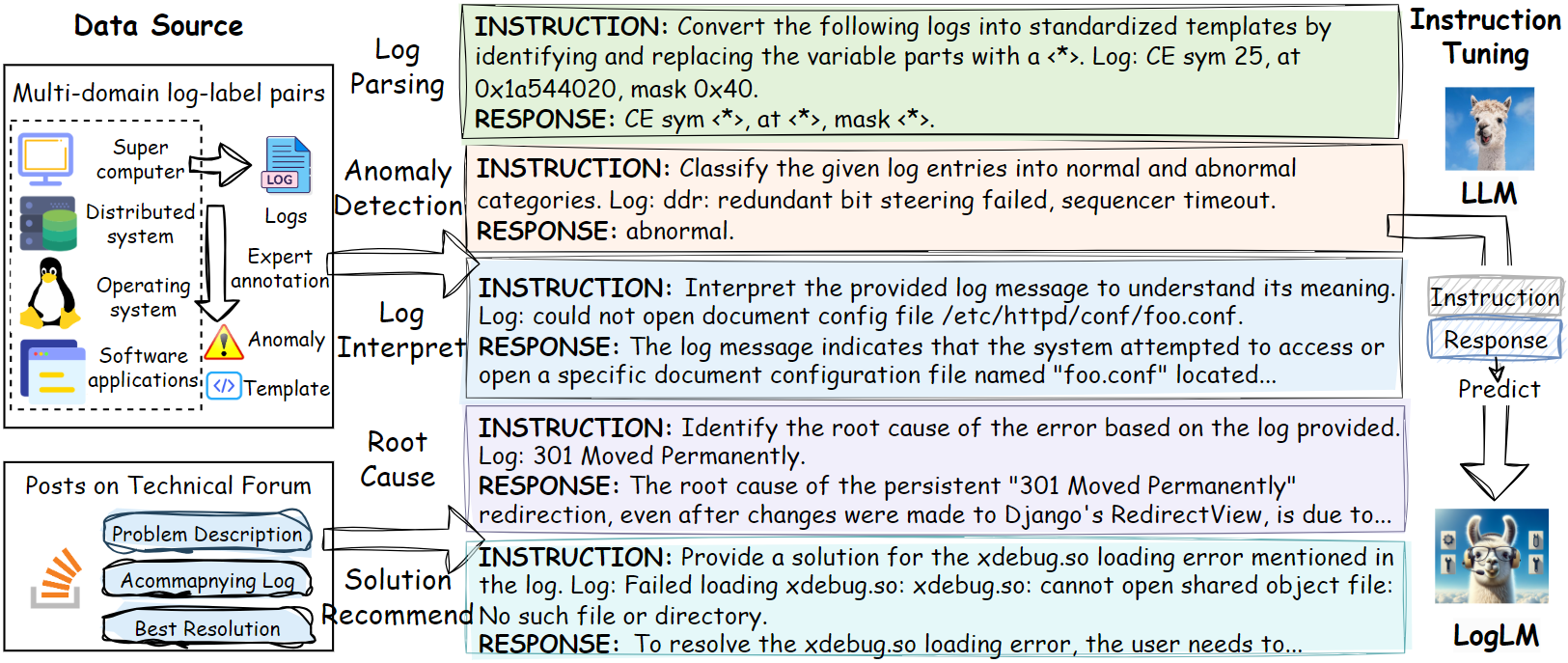}
  \caption{Illustration on the capabilities composition, training dataset construction and training of LogLM.}
  \label{fig_method}
\end{figure*}

\subsubsection{Other Analysis Tasks}

In addition to the two widely investigated tasks, several other tasks in log analysis are also studied. Meng~\emph{et al.}~\cite{meng2023logsummary} proposed the task of log summary, which involves summarizing key phrases from log sequences. Liu~\emph{et al.}\cite{liu2024logprompt} proposed log interpretation, a task that aims to explain key elements in logs and describe the logs in natural language. Chen~\emph{et al.}~\cite{chen2024automatic} implemented an on-call system empowered by LLMs that predicts cloud incidents' root cause category. Ahmed~\emph{et al.}~\cite{ahmed2023recommending} utilized fine-tuned LLMs to handle cloud incidents and recommend mitigation steps for engineers.

\subsubsection{LogLM \emph{v.s.} Existing Approaches}

At earlier times there were endeavors which attempted to unify multiple log analysis tasks, such as Unilog~\cite{zhu2021unilog} and Biglog\cite{tao2023biglog}, where a universal log encoder was trained for performing multiple downstream tasks. However, they still require fine-tuning on task-specific and domain-specific log-label pairs to update the parameters in classification layers of the model, leading to multiple different models for varied downstream tasks in the deployment. Also, they can't follow complex user instructions or perform unseen tasks. LogPrompt~\cite{liu2024interpretable} is another method supporting both log parsing and anomaly detection. But it requires meticulous prompt engineering for each task. And the underlying LLM is API-based ChatGPT, which limits the application in industry.

Compared with existing approaches, LogLM unifies multiple log tasks and is trained from open-source LLMs, thereby distinguishing by its potentials of single-model deployment for multi-task analysis in industrial applications. Moreover, LogLM can follow complex user instructions and perform analysis tasks that are unseen from training set, increasing the flexibility and intelligence of automated software O\&M.

\section{Methodology}

Fig.~\ref{fig_method} provides an overview of our methodology to build LogLM. We first define five closely interconnected log analysis capabilities for LogLM, which facilitate a profound comprehension of log data and a robust application of log analysis. To construct a high-quality instruction dataset, we identified publicly available data sources comprising real-world logs and user posts from technical communities. We designed an LLM-assisted workflow to transform log-label pairs and log-problem-resolution triplets from these data sources into the format of (\textsc{Instruction}, \textsc{Response}), covering all five analysis capabilities. The instruction pairs are then utilized to fine-tune LogLM from foundation LLMs.   

\subsection{Analysis Capabilities Composition}
To facilitate a comprehensive log-related analytical capacity for LogLM, it is imperative to meticulously select the types of instructions utilized during the instruction tuning process. Based on existing studies in the field of log analysis, five most representative log analysis capabilities are considered in composition of the instruction dataset of LogLM. These five capabilities are not discrete entities; rather, they are closely interconnected, thereby fostering a deeper understanding of log-related analytical tasks by the model. Specifically, these capabilities can be categorized into two distinct groups.

\textbf{(1) Log Understanding Capabilities}, which consists of structural understanding (\textit{Log Parsing}) and semantic understanding (\textit{Log Interpretation}). \textit{Log Parsing} involves decomposing a log's structure into its template and variable components, which aids in comprehending the overall structure of logs. Conversely, \textit{Log Interpretation} expands the semantic content of logs into coherent natural language sentences, thereby enhancing the understanding of log semantics. A precise decomposition of log structure and an accurate interpretation of log semantics is the foundation of subsequent applications.

\textbf{(2) Log Application Capabilities}, including three log-related application tasks: \textit{Anomaly Detection}, \textit{Root Cause Analysis} and \textit{Solution Recommendation}. These three capabilities follow a logical sequence that reflects the operational procedures typically employed by O\&M engineers during log analysis. When system anomaly is detected from logs, engineers need to find the cause of the breakdown and take measures to solve the problem. Thus, the integration of these three application capabilities, along with the two understanding capabilities, addresses the practical requirements encountered in real-world scenarios, thereby enhancing the overall performance of log analysis.    

In real-world practice, the five capabilities are indispensable for an engineer to perform log analysis. For example, Tom, a web O\&M engineer, spotted a server log: ``2024-12-25 08:32:00, ERROR 500 /checkout - Database Timeout''. By filtering out irrelevant variables (through \textit{Log Parsing}), Tom focused on ``ERROR 500'' and ``Database Timeout''. Based on his expertise (through \textit{Log Interpretation}), this log was interpreted by Tom as a failed database connection, which suggested a potential system anomaly (\textit{Anomaly Detection}). By examining server resources, Tom found the root cause of this incident: the recent database update had introduced a change in query execution plans, which overloaded the server during peak traffic (\textit{Root Cause Analysis}). Tom then recommended several solutions in a group meeting, including rolling back the update, scaling up the database server and setting up proactive alerts (\textit{Solution Recommendation}). \textbf{By including these capabilities into the training set, the LLM can comprehensively learn essential skills for log analysis and maximally aligns with real-world human engineers.}  
\subsection{Instruction Dataset Construction}
Upon determining the capabilities composition, we seek to construct pairs of (\textsc{Instruction},\textsc{Response}) within these capabilities. Log parsing and anomaly detection are two well-established tasks, thereby the instruction pairs of the two capabilities were constructed from log-label pairs in representative open-source datasets. For the other three capabilities, it is challenging to find off-the-shelf open-source datasets. Thus, we designed an LLM-assisted workflow to decompose log-related high-quality community posts on technical forums, into instruction pairs of the three capabilities. The statistics of the constructed instruction dataset for training is shown in Table~\ref{tab:traing_set}, and the specific construction process is discussed below.
\begin{table}[tbp]
\caption{Statistics of the Instruction Dataset for Training LogLM}
% \begin{center}
\centering
\resizebox{\linewidth}{!} {%
\begin{tabular}{p{0.25\linewidth}@{\hskip 0.01in}c@{\hskip 0.01in}c@{\hskip 0.01in}c}
\toprule
\multicolumn{1}{l}{\textbf{Capabilities}} & \multicolumn{1}{l}{\textbf{Data Source}}& \begin{tabular}[c]{@{}l@{}} \multicolumn{1}{c}{\textbf{\# Instruction}} \\ \multicolumn{1}{c}{\textbf{Pairs}} \end{tabular} & \multicolumn{1}{l}{\textbf{Domain}}\\
\midrule
\multicolumn{1}{l}{\multirow{7}{*}{Log Parsing}} & \multirow{7}{*}{Loghub2k~\cite{he2020loghub}} & 200 & HDFS\\
 &  & 200 & Hadoop\\
 &  & 200 & Zookeeper\\
 &  & 200 & BGL\\
 &  & 200 & HPC\\
 &  & 200 & Linux\\
 &  & 200 & Proxifier\\
 \midrule
\multicolumn{1}{l}{\multirow{2}{*}{Anomaly Detection}} & \multirow{2}{*}{LogPrompt~\cite{liu2024interpretable}} & 194 & BGL\\
 &  & 138 & Spirit\\
 \midrule
\multicolumn{1}{l}{Log Interpretation}& \multirow{3}{*}{LogExpert~\cite{wang2024logexpert}} & 300 & \multirow{3}{*}{Apache}\\
\multicolumn{1}{l}{Root Cause Analysis}&  & 300 & \\
\multicolumn{1}{l}{Solution Recommendation}&  & 300 & \\
\hdashline
\noalign{\vskip 4pt}
\multicolumn{1}{l}{All}& - & 2632 & -\\
\bottomrule
\end{tabular}
}
\label{tab:traing_set}
% \end{center}
\end{table}

\subsubsection{Log Parsing}\label{sec:method_parsing}

As shown in Table~\ref{tab:traing_set}, the source data utilized for constructing instruction pairs related to \textit{Log Parsing} is the Loghub2k dataset~\cite{he2020loghub}, which comprises real-world logs derived from a variety of domains. Zhu \emph{et al.}~\cite{zhu2019tools} manually extracted log templates from a subset of 2000 logs across multiple domains, thereby establishing a benchmark that is considered most representative for the log parsing task. Consistent with prior research~\cite{tao2023parser,tao2023biglog,liu2023multi}, log-template pairs from seven prominent domains—including supercomputers, distributed systems, operating systems, and software applications—were selected to formulate instruction pairs.

As discussed in Section~\ref{sec:intro}, our objective is to mitigate reliance on in-domain and task-specific log data. Thus, we incorporated only a limited selection of log-template pairs into the instruction dataset, reserving the majority for testing purposes. This approach effectively simulates real-world online scenarios where in-domain log data for training is often scarce~\cite{tao2022logstamp}. Specifically, we adhered to the setting in LogPrompt~\cite{liu2024interpretable}, wherein the chronologically first 10\% of logs were selected from each domain to construct the training set and the remaining 90\% were for evaluation. This yielded a total of 1400 log-template pairs across the seven domains.

In constructing instruction pairs for \textit{Log Parsing}, as shown in Fig.~\ref{fig_method}, we employed a straightforward prompt proposed by Liu \emph{et al.}~\cite{liu2024interpretable} as \textsc{Instruction}. This prompt has demonstrated consistent parsing performance when applied to ChatGPT, thereby qualifying it as a high-quality \textsc{Instruction}. For the \textsc{Response}, the manually extracted templates from the dataset provide a human-preferred answer to the \textsc{Instruction}, thereby enhancing human alignment in subsequent instruction tuning processes.

\subsubsection{Anomaly Detection}\label{sec:method_anomaly}

Instruction pairs related to \textit{Anomaly Detection} were derived from the BGL and Spirit benchmark datasets, which were meticulously curated by Oliner~\emph{et al.}\cite{oliner2007supercomputers}. These datasets comprise system logs collected from supercomputers, with annotations performed in collaboration with domain experts to identify anomalous events for the anomaly detection task. Liu~\emph{et al.}\cite{liu2024interpretable} extracted a total of 1766 log templates from the BGL dataset and 1297 from the Spirit dataset, and released these pairs of log templates and anomaly labels. From this collection, we randomly selected a minor subset—approximately 10\%—from each dataset to construct the instruction pairs, while reserving the remaining data for evaluation purposes. To ensure a balanced representation of abnormal and normal samples, we ensured that each subset contained around 10\% abnormal samples, thereby reflecting the original distribution of abnormal and normal log messages in the source data. For the construction of the instruction pairs, as shown in Fig.~\ref{fig_method}, we employed a straightforward prompt for the anomaly detection task proposed by Liu~\emph{et al.}~\cite{liu2024interpretable}, with a single log as an input and the normal/abnormal label serving directly as the \textsc{Response}. For new logs, multiple entries can be concatenated and input as a single log.

\subsubsection{Interpretation, Root Cause \& Solution (IRS)}\label{sec:method_IRS}

For IRS capabilities, the availability of off-the-shelf datasets that encompass logs and their corresponding task outputs is limited. However, community contributions on technical forums may offer valuable data sources useful for the development of IRS capabilities for LogLM. Specifically, when users post technical issues accompanied by error logs generated by software, their descriptions often contain the direct causes of the errors present in the logs, as well as the contextual circumstances surrounding those logs, which help in interpreting the logs. Furthermore, the highest-voted recommended resolutions typically include interpretations of the logs, identifications of the root causes of the errors, and user-preferred solutions.

To achieve this, we utilize a dataset released by Wang \emph{et al.}~\cite{wang2024logexpert}, which consists of log-related user posts from Stack Overflow tagged with "apache". This dataset was meticulously curated to include only those user posts that contained at least one accompanying log. Additionally, the authors manually verified the feasibility of the highest-voted resolutions for each post, resulting in a high-quality dataset comprising 384 log-problem-resolution triplets. Subsequently, we composed a prompt to instruct GPT-4 to decompose three instruction pairs, regarding to \textit{Log Interpretation}, \textit{Root Cause Analysis}, and \textit{Solution Recommendation}, from each log-problem-resolution triplet within the dataset. The specific prompt utilized is:

\begin{mybox}
Assume you are a dedicated IT specialist focusing on log analysis for system O\&M. The following real-world case includes a log, a title, a description of related user posted problem and a community solution. Your task is to decompose three INSTRUCTION-INPUT-RESPONSE pairs based on the real-world case. Requirement for INSTRUCTION: The topic of three instructions are interpretation of the log, root cause of the log and solution of the log, respectively. Organize each instruction to be a concise user query on the log. Requirement for INPUT: The three inputs are all the given log in the real-world case. If the original log is quite long, retain only necessary part. Requirement for RESPONSE: The three responses must properly meet the instructions. Organize your language to be precise and professional. Format your answer like this: INSTRUCTION 1: xxx\verb|\n| INPUT 1: xxx\verb|\n| RESPONSE 1: xxx\verb|\n| INSTRUCTION 2: xxx\verb|\n| INPUT 2: xxx\verb|\n| RESPONSE 2: xxx\verb|\n| INSTRUCTION 3: xxx\verb|\n |INPUT 3: xxx\verb|\n| RESPONSE 3: xxx. The case begins: \textcolor{red}{\{Input\}}
\end{mybox}

By applying this prompt to GPT-4 (model version: \textit{gpt-4-turbo-preview}), we generated 384 instruction pairs corresponding to each capability, with \textsc{Instruction} reflecting the specific requests merged with the input log, and \textsc{Response} being derived from human-authored posts. To mitigate the risk of generating inaccurate or irrelevant content, we implemented a human calibration process to ensure that the generated instruction pairs were consistent with the original user posts and expert resolutions. Approximately 2\% of the generated content was excluded following this inspection, resulting in 376 calibrated instruction pairs for each of the three capabilities. These instruction pairs were subsequently divided into training and testing sets in an 8:2 ratio (\emph{i.e.}, 300:76).
\subsection{Training of LogLM}

The instruction dataset in Table~\ref{tab:traing_set} serves as a paradigm for properly responding to human instructions in log analysis, which the LLM learns to emulate during instruction tuning. Starting with a pre-trained foundation model, the LLM gradually aligns its pre-existing knowledge with human-preferred analysis outputs by predicting the next tokens in the \textsc{Response} conditioned on an input \textsc{Instruction}. 

Denote the instruction pair by $x=($\textsc{Instruction}, \textsc{Response}$)$, where \( x \in C \) represents elements in the constructed instruction dataset \( C \). The parameters of the foundation model are denoted as \( \theta \), and after instruction tuning, the parameters are updated to \( \theta_c \), which represents the adapted model, LogLM. The goal is to maximize the log likelihood of next tokens (\emph{i.e.}, minimal unit of a tokenized sentence) in the \textsc{Response} conditioned on the corresponding \textsc{Instruction}. By denoting the tokens in the \textsc{Response} of $x$ as $( y_1, y_2, \ldots, y_{N_x})$, where \( N_x \) is the number of tokens in the \textsc{Response}, the training objective can be formulated as:

\begin{equation}\label{eq1}
   \theta_c = \argmax_{\theta} \sum\limits_{x \in C} \sum\limits_{i=1}^{N_x} \log P(y_i \,|\, \textsc{Instruction}, y_{1:i-1}; \theta, x).
\end{equation}

\section{Experiment}\label{sec:exp}
\subsection{Implementation Details}

In the experiment, the foundation model $\theta$ is LLaMA-2-7B~\cite{touvron2023llama2}, an open-source LLM with 7B parameters. Our main implementation of LogLM, denoted by LogLM-7B, was trained on the full instruction dataset in Table~\ref{tab:traing_set} according to Eq.~\eqref{eq1}. During instruction tuning, the learning rate is $2\times10^{-5}$, the batch size is 32 and the number of training epochs is six. Other parameters follow the default settings in LLaMAFactory~\cite{zheng-etal-2024-llamafactory}, the framework we utilized for LLM training and inferencing.

\subsection{Research Questions \& Key Findings}

In this section, we describe the research questions (RQ) on evaluating LogLM and the key findings in experiments.

\textbf{RQ1:} Can the instruction-based approach outperform task-based methods in the five log analysis capabilities?

\textbf{Key Findings of RQ1:} To assess LogLM's log analysis capabilities, in Section~\ref{sec:RQ1}, we compare its performance with existing task-based methods across five benchmarks representing different log analysis capabilities. As a single model, LogLM-7B consistently outperforms 20 existing approaches across different analysis tasks, highlighting its strong industrial application potential due to its cost-effectiveness.

\textbf{RQ2:} (A) What contributes to LogLM's superior performance? (B) Can LogLM reduce reliance on in-domain and task-specific log data? (C) Does the model benefit from increased diversity and quantity of training instructions?

\textbf{Key Findings of RQ2:} Section~\ref{sec:RQ2} addresses the three sub-questions through an ablation study on the training data composition of LogLM. Beginning with merely in-domain training data for a specific task, we observe a substantial improvement in LogLM's average performance as instruction pairs from other domains and tasks are introduced. This demonstrates that LogLM’s advantage stems from learning connections between instruction pairs across diverse domains and tasks (RQ2-A). Furthermore, this supports answering RQ2-B, as the majority of LogLM’s training data come from other domains and tasks, reducing the need for extensive data collection in a private domain or for an emerging task in industrial applications. Regarding RQ2-C, the finding suggests that training on instruction pairs from a diverse set of log domains and analysis tasks enhances LogLM's performance, highlighting its application potential in complex systems.

\textbf{RQ3:} Can LogLM generalize to unseen tasks and follow complex user instructions?

\textbf{Key Findings of RQ3:} In Section~\ref{sec:unseen_task}, we train LogLM using instruction pairs from only four capabilities, leaving the remaining one as an unseen task. Despite this, LogLM achieved best performance on 10 of 12 test items in these unseen tasks. Additionally, a case study in Section~\ref{sec:complex_instruction} demonstrate LogLM-7B's ability to answer knowledge-based questions related to log analysis, handle previously unseen instructions, and perform a combination of tasks.

The rest of Section~\ref{sec:exp} is organized as follows. Section~\ref{sec:RQ1} addresses RQ1, evaluating \textit{Log Parsing} in Section~\ref{sec:RQ1_parsing}, \textit{Anomaly Detection} in Section~\ref{sec:RQ1_anomaly} and the other three capabilities in Section~\ref{sec:RQ1_IRS}. Section~\ref{sec:RQ2} and \ref{sec:RQ3} address RQ2 and RQ3, respectively.
\begin{table*}[t!]
\caption{Benchmarking on the Capability of Log Parsing Evaluated in both Coarse-level and Fine-level}
\centering
\resizebox{0.9\linewidth}{!} {%
\setlength{\tabcolsep}{2pt}
\begin{tabular}{l@{\hskip 0.1in}c@{\hskip 0.05in}c@{\hskip 0.1in}c@{\hskip 0.05in}c@{\hskip 0.1in}c@{\hskip 0.05in}c@{\hskip 0.1in}c@{\hskip 0.05in}c@{\hskip 0.1in}c@{\hskip 0.05in}c@{\hskip 0.1in}c@{\hskip 0.05in}c@{\hskip 0.1in}c@{\hskip 0.05in}c@{\hskip 0.1in}||c@{\hskip 0.1in}c@{\hskip 0.05in}c}
%\begin{tabular}{lcccccccccccccc||ccc}
\toprule
\multirow{2}{*}{{\textbf{Methods}}} & \multicolumn{2}{>{\hspace{-1em}\centering}c}{\textbf{HDFS}} & \multicolumn{2}{>{\hspace{-1em}\centering}c}{\textbf{Hadoop}} & \multicolumn{2}{>{\hspace{-1em}\centering}c}{\textbf{Zookeeper}} & \multicolumn{2}{>{\hspace{-1em}\centering}c}{\textbf{BGL}} & \multicolumn{2}{>{\hspace{-1em}\centering}c}{\textbf{HPC}} & \multicolumn{2}{>{\hspace{-1em}\centering}c}{\textbf{Linux}} & \multicolumn{2}{>{\hspace{-1em}\centering}c}{\textbf{Proxifier}} & \multicolumn{2}{c}{\textbf{Avg.}} \\ \cmidrule(l{-0.3em}r{0.8em}){2-3} \cmidrule(l{-0.3em}r{0.8em}){4-5} \cmidrule(l{-0.3em}r{0.8em}){6-7} \cmidrule(l{-0.3em}r{0.8em}){8-9} \cmidrule(l{-0.3em}r{0.8em}){10-11} \cmidrule(l{-0.3em}r{0.8em}){12-13} \cmidrule(l{-0.3em}r{0.8em}){14-15} \cmidrule(lr){16-17}
& \textbf{RI$^{\mathrm{a}}$} & \textbf{F1} & \textbf{RI} & \textbf{F1} & \textbf{RI} & \textbf{F1} & \textbf{RI} & \textbf{F1} & \textbf{RI} & \textbf{F1} & \textbf{RI} & \textbf{F1} & \textbf{RI} & \textbf{F1} & \textbf{RI} & \textbf{F1}\\
\midrule
\addlinespace
IPLoM \cite{makanju2009clustering} & 0.914 & 0.389 & 0.636 & 0.068 & 0.787 & 0.225 & 0.858 & 0.391 & 0.228 & 0.002 & 0.695 & 0.225 & 0.822 & 0.500 & 0.706 & 0.257 \\
LKE~\cite{fu2009execution} & 0.861 & 0.424 & 0.150 & 0.198 & 0.787 & 0.225 & 0.848 & 0.379 & 0.119 & 0.381 & 0.825 & 0.388 & 0.379 & 0.309 & 0.567 & 0.329 \\
LogSig~\cite{tang2011logsig} & 0.872 & 0.344 & 0.651 & 0.050 & 0.787 & 0.225 & 0.806 & 0.333 & 0.119 & 0.002 & 0.715 & 0.146 & 0.559 & 0.339 & 0.644 & 0.206 \\
FT-tree~\cite{zhang2017syslog} & 0.908 & 0.385 & 0.668 & 0.046 & 0.773 & 0.186 & 0.275 & 0.497 & 0.119 & 0.002 & 0.709 & 0.211 & 0.722 & 0.420 & 0.596 & 0.250 \\
Spell~\cite{du2016spell} & 0.871 & 0.000 & 0.721 & 0.058 & 0.102 & 0.045 & 0.503 & 0.536 & 0.882 & 0.000 & 0.706 & 0.091 & 0.621 & 0.000 & 0.629 & 0.104 \\
Drain~\cite{he2017drain} & 0.914 & 0.389 & 0.647 & 0.068 & 0.787 & 0.225 & 0.822 & 0.397 & 0.119 & 0.002 & 0.695 & 0.225 & 0.822 & 0.500 & 0.687 & 0.258 \\
MoLFI~\cite{messaoudi2018search} & 0.871 & 0.000 & 0.699 & 0.095 & 0.899 & 0.000 & 0.792 & 0.333 & 0.881 & 0.000 & 0.410 & 0.026 & 0.621 & 0.000 & 0.739 & 0.065 \\
LogParse~\cite{meng2020logparse} & 0.907 & 0.632 & 0.349 & 0.502 & 0.982 & 0.348 & 0.992 & 0.665 & 0.194 & 0.330 & 0.825 & 0.588 & 0.490 & 0.334 & 0.677 & 0.486 \\
LogStamp~\cite{tao2022logstamp} & 0.954 & 0.523 & 0.927 & 0.594 & 0.992 & 0.275 & 0.984 & 0.818 & 0.949 & 0.434 & 0.760 & 0.658 & 0.811 & 0.438 & 0.911 & 0.534 \\
%LogPPT \cite{le2023log} & 0.960 & 0.838 & 0.987 & 0.526 & 0.988 & 0.795 & 0.859 & \textbf{0.982} & 0.238 & 0.287 & 0.831 & 0.423 & 0.804 & 0.638 & 0.810 & 0.641 \\
LogPrompt~\cite{liu2024logprompt} & 0.890 & 0.863 & 0.879 & 0.763 & 0.948 & 0.889 & 0.964 & 0.865 & 0.934 & 0.759 & 0.758 & 0.766 & 0.567 & 0.653 & 0.849 & 0.794 \\
\hdashline
\noalign{\vskip 2pt}
\textbf{LogLM-7B} & \textbf{1.000} & \textbf{0.998} & \textbf{1.000} & \textbf{0.973} & \textbf{1.000} & \textbf{0.995} & \textbf{0.999} & \textbf{0.977} & \textbf{0.999} & \textbf{0.935} & \textbf{0.994} & \textbf{0.934} & \textbf{0.879} & \textbf{0.940} & \textbf{0.982} & \textbf{0.965} \\
\bottomrule
\multicolumn{17}{l}{$^{\mathrm{a}}$ \textbf{RI} stands for coarse-level RandIndex. \textbf{F1} stands for fine-level F1-score.} \\
\end{tabular}
}
\label{tab:logParsing_exp}
\end{table*}
\subsection{RQ1: Benchmarking on Log Analysis Capabilities}\label{sec:RQ1}

\subsubsection{Log Parsing}\label{sec:RQ1_parsing}

\paragraph{Evaluation Setting}\label{sec:logParsing_exp_setting}

As is discussed in Section~\ref{sec:method_parsing}, this benchmark involves testing the performance of log parsing on the chronologically last 90\% of the logs from seven domains in the Loghub2k~\cite{he2020loghub} dataset. We compare LogLM with 10 existing approaches specifically designed for the task of log parsing, encompassing cluster-based methods~\cite{tang2011logsig,fu2009execution}, heuristic methods~\cite{messaoudi2018search, du2016spell,makanju2009clustering}, tree-based methods~\cite{zhang2017syslog,he2017drain}, machine learning methods~\cite{meng2020logparse} and LLM-based methods~\cite{tao2022logstamp, liu2024logprompt}. Following the experiment setting in Liu~\emph{et al.}~\cite{liu2024interpretable}, each baseline is trained on the first 10\% of the logs in each domain, except for LogPrompt~\cite{liu2024logprompt}, which directly utilize ChatGPT for log parsing without training. In other words, for each domain, LogLM and other approaches only see the first 10\% of logs in the specific domain during training and the rest 90\% of logs are unseen, thereby simulating online situations while ensuring a fair comparison.

Following Liu~\emph{et al.}~\cite{liu2024interpretable}, two most popular evaluation metrics for the task of log parsing among existing studies~\cite{meng2020logparse,tao2023biglog,tao2022logstamp} are utilized, encompassing both coarse-level (RandIndex~\cite{rand1971objective}) and fine-level (F1-score). RandIndex assesses the accuracy of log clustering (\emph{i.e.}, whether two logs with the same template are accurately clustered together), regardless of the correctness of variables in the extracted templates. In contrast, F1-score measures the precise identification of variable parts in logs, thereby serving as a fine-level metric. To calculate the F1-score,  the predicted log template is tokenized into a list of tokens. Then, for each token, count the terms $TP$, $TN$, $FP$, and $FN$. If the token is truly a variable and is correctly predicted as such (or not), increment the $TP$ (or $FP$) by one. If the token is not a variable and is predicted not to be a variable (or be), increment the $TN$ (or $FN$) by one. The F1-score is computed as the harmonic average of Recall ($Recall = \frac{TP}{TP+FN}$) and Precision ($Precision = \frac{TP}{TP+FP}$).

\paragraph{Result}

As shown in Table~\ref{tab:logParsing_exp}, our model, LogLM-7B, exhibited remarkable performances on this benchmark, outperforming existing approaches significantly in both coarse-level evaluation and fine-level evaluation. Averagely, LogLM-7B outperforms best baselines by 7.79\% in RandIndex and by 10.03\% in F1-score. This advantages indicate that LogLM can accurately recognize variable parts in logs while extracting correct coarse-level templates, even when the in-domain training logs are limited in quantity.  

\begin{table}[t!]
    \caption{Benchmarking on the capability of Anomaly Detection}
    \centering
    \resizebox{0.84\linewidth}{!} {%
    \begin{tabular}{l@{\hskip 0.15in}c@{\hskip 0.1in}c@{\hskip 0.1in}c@{\hskip 0.15in}c@{\hskip 0.1in}c@{\hskip 0.1in}c}
    \toprule
    \multirow{2}{*}{\textbf{Methods}} & \multicolumn{3}{c}{\hspace{-1.5em}\textbf{BGL}}       & \multicolumn{3}{c}{\hspace{-0.2em}\textbf{Spirit}} \\ \cmidrule(l{0em}r{1.5em}){2-4} \cmidrule(l{-0.5em}r{0.2em}){5-7} 
                                   & \hspace{0.6em}\textbf{S-F1}$^{\mathrm{a}}$ & \textbf{T-F1} &   & \textbf{S-F1} & \textbf{T-F1} &    \\ \midrule
    DeepLog~\cite{du2017deeplog}               & 0.194             & -         &    & 0.092             & -         &    \\
    LogAnomaly~\cite{meng2019loganomaly}        & 0.129             & -        &    & 0.138             & -         &    \\
    LogRobust~\cite{zhang2019robust}            & 0.536             & -         &    & 0.045             & -         &    \\
    ChatGPT~\cite{liu2024interpretable}            & 0.129             & 0.067          &    & 0.122             & 0.050          &    \\
    LogPrompt~\cite{liu2024interpretable}          & 0.314             & 0.233          &    & 0.144             & 0.071          &    \\
    \hdashline
    \noalign{\vskip 2pt}
    \textbf{LogLM-7B}        & \textbf{0.811}    & \textbf{0.625}          &    & \textbf{0.584}    & \textbf{0.278}          &    \\ \bottomrule 

    \multicolumn{7}{l}{$^{\mathrm{a}}$ \textbf{S-F1}/\textbf{T-F1} means F1-Score in session/template-level.} \\
    \end{tabular}
    }
    \label{tab:anomaly_exp}
\end{table}
\subsubsection{Anomaly Detection}\label{sec:RQ1_anomaly}

\begin{table*}[t!]
\caption{Benchmarking with Existing LLMs on Log Interpretation, Root Cause Analysis, and Solution Generation}
\centering
\resizebox{0.84\linewidth}{!}{
\begin{tabular}{@{}l@{\hskip 0.1in}c@{\hskip 0.05in}c@{\hskip 0.05in}c@{\hskip 0.05in}c@{\hskip 0.1in}c@{\hskip 0.05in}c@{\hskip 0.05in}c@{\hskip 0.05in}c@{\hskip 0.1in}c@{\hskip 0.05in}c@{\hskip 0.05in}c@{\hskip 0.05in}c@{}}
\toprule
\multirow{2}{*}{\textbf{Methods}} & \multicolumn{4}{c}{\hspace{-0.4em}\textbf{Log Interpretation}}                              & \multicolumn{4}{c}{\hspace{-1.0em}\textbf{Root Cause Analysis}}                                   & \multicolumn{4}{c}{\hspace{-0.2em}\textbf{Solution Recommendation}}                                   \\ \cmidrule(l{0.3em}r{1.2em}){2-5}  \cmidrule(l{0em}r{1.2em}){6-9} \cmidrule(l{0em}){10-13} 
                                  & \hspace{0.3em}\textbf{BLEU}   & \hspace{0.5em}\textbf{R-1}$^{\mathrm{a}}$ & \textbf{R-2} & \textbf{R-L} & \textbf{BLEU}   & \textbf{R-1} & \textbf{R-2} & \textbf{R-L} & \textbf{BLEU}  & \textbf{R-1} & \textbf{R-2} & \textbf{R-L} \\ \midrule
LLaMA-3-70B~\cite{dubey2024llama}              & 0.507           & 7.984            & 2.121            & 5.864            & 0.172           & 6.876            & 1.273            & 4.382            & 0.529          & 7.640            & 1.557            & 5.368            \\
LLaMA-3.1-405B~\cite{dubey2024llama}           & 4.466           & 28.563           & 11.106           & 17.071           & 1.416           & 12.984           & 3.694            & 8.162            & 2.018          & 18.768           & 5.287            & 12.045           \\
Claude-3.5-Sonnet~\cite{claude}        & 4.324           & 30.512           & 11.786           & 18.307           & 2.563           & 25.824           & 7.909            & 15.221           & 2.698          & 23.618           & 6.936            & 15.109           \\
GPT-4~\cite{achiam2023gpt}                    & 2.831           & 25.205           & 9.147            & 13.683           & 2.057           & 21.810           & 6.544            & 11.518           & 1.502          & 21.964           & 5.532            & 12.274           \\
OWL-7B~\cite{guoowl}                   & 2.566           & 28.211           & 8.289            & 19.123           & 1.947           & 20.893           & 5.671            & 14.718           & 0.953          & 21.620           & 5.006            & 15.574           \\
\hdashline
\noalign{\vskip 2pt}
\textbf{LogLM-7B}              & \textbf{15.584} & \textbf{46.488}  & \textbf{23.087}  & \textbf{34.769}  & \textbf{12.398} & \textbf{40.602}  & \textbf{19.042}  & \textbf{30.227}  & \textbf{8.241} & \textbf{34.415}  & \textbf{13.911}  & \textbf{25.431}  \\ \bottomrule 
\multicolumn{13}{l}{$^{\mathrm{a}}$ \textbf{R-1} stands for ROUGE-1. \textbf{R-2} stands for ROUGE-2. \textbf{R-L} stands for ROUGE-L.}\\
\end{tabular}
}\label{tab:IRS_exp}
\end{table*}
\paragraph{Evaluation Setting}\label{sec:anomaly_exp_setting}

This evaluation involves comparing LogLM with both template-level methods~\cite{liu2024interpretable} and session-level methods~\cite{du2017deeplog,meng2019loganomaly,zhang2019robust}, as discussed in Section~\ref{sec:anomaly_relatedwork}. Thus, the evaluation settings are split into template-level and session-level. For template-level, the test set is the split template-label pairs as discussed in Section~\ref{sec:method_anomaly}, containing around 90\% of the templates extracted by Liu~\emph{et al.}~\cite{liu2024interpretable} from the dataset of BGL and Spirit. For session-level, following the setting in LogPrompt, log sessions were built using fixed-window grouping with a length of 100 chronologically adjacent logs in BGL and Spirit. The first 4000 logs in each dataset are used for training baselines and the rest logs are for testing. To avoid data leakage, logs in the training set of LogLM were excluded from the session-level test set, leading to a final test set of 40521 and 7515 sessions for BGL and Spirit. 

The evaluation metric for both the template-level and session-level is the F1-score of anomaly. F1-score in template-level (or session-level) considers accurate recall of abnormal logs (or sessions) from test cases and the precise prediction of anomaly in template-level (or session-level). Similar to Section~\ref{sec:logParsing_exp_setting}, $TP$ represents the successful identification of an anomaly (similarly for $TN$, $FP$, and $FN$) and the F1-score is calculated as the harmonic average of Recall and Precision.

\paragraph{Result}

The result is shown in Table~\ref{tab:anomaly_exp}. As is observed in the study of Liu~\emph{et al.}~\cite{liu2024interpretable}, detecting anomaly in online situations with limited number of training logs can be challenging for most methods, including even ChatGPT. Normally, without fitting on enough historical data, it is hard to precisely model the anomaly patterns in domain-specific logs. However, by training on diverse log tasks, LogLM possesses more knowledge in log analysis, thereby can model in-domain anomaly patterns more efficiently. As a result, LogLM-7B achieves a strong result in Table~\ref{tab:anomaly_exp} both for BGL and Spirit, while utilizing only 100+ in-domain logs in training.

\subsubsection{Interpretation, Root Cause \& Solution (IRS) Tasks}\label{sec:RQ1_IRS}

\paragraph{Evaluation Setting}

The evaluation sets for IRS tasks encompass 76 Q\&A pairs for each capability, constructed from user posts and highest-voted resolutions in real-world technical community as described in Section~\ref{sec:method_IRS}. For baselines, general-purpose LLMs are involved given their strong abilities of solving complex problems~\cite{kalyan2023survey,kung2023performance}, including LLaMA-3-70B~\cite{dubey2024llama}, LLaMA-3.1-405B~\cite{dubey2024llama}, Claude-3.5-Sonnet~\cite{claude} and GPT-4~\cite{achiam2023gpt}. In addition, OWL-7B~\cite{guoowl}, a domain-purpose LLM designed for IT operations, is also included in the evaluation. All baselines are directly prompted with instructions in the evaluation sets (\emph{i.e.}, zero-shot). For fairness, we examine the zero-shot (\emph{i.e.}, no in-domain training data) performance of LogLM in Section~\ref{sec:unseen_task}.

The evaluation employs four commonly-used metrics in Q\&A tasks for measuring the accuracy of generated answers compared with reference answers. ROUGE-1 and ROUGE-2~\cite{lin2004rouge} assess the overlap of unigrams and bigrams (\emph{i.e.}, a sequence of one or two consecutive words), respectively, between the generated and reference texts, focusing on successful recalls of information in reference answers. ROUGE-L~\cite{lin2004rouge} measures the longest common subsequence (LCS) between the two texts, capturing sentence-level structural similarity. BLEU~\cite{papineni2002bleu} evaluates precision of generated answers by counting percentage of n-gram (\emph{i.e.}, n consecutive words) overlap in the generated texts, penalizing brevity and rewarding fluency.

Each of these metrics ranges from 0 to 100. A score of 100 indicates a perfect match between the generated and reference text, while a score of 0 indicates no overlap. ROUGE-1 and ROUGE-2 are computed as the ratio of overlapping n-grams in the generated text to number of n-grams in the reference text. ROUGE-L is calculated as the length of LCS to the length of the reference text. BLEU, however, is calculated as the geometric mean of n-gram precision (\emph{i.e.}, overlapping n-grams to number of n-grams in the predicted text) from $n=1$ to $n=4$, with a brevity penalty to penalize short outputs. These four metrics capture both surface-level lexical overlap and deeper structural similarities, which are essential for Q\&A evaluation.

\begin{figure*}[t!]
    \centering
  \includegraphics[width=\linewidth]{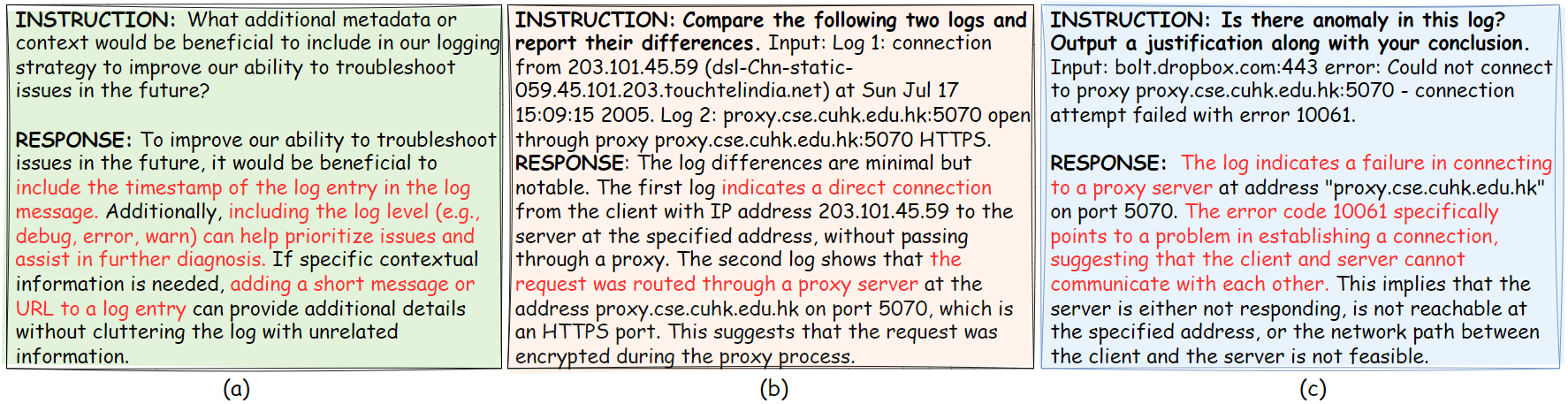}
  \caption{Three cases of LogLM-7B responding to complex user instructions: (a) log-related Q\&A, (b) unseen new task, and (c) combination of tasks.}
  \label{fig_user_instruction}
\end{figure*}

\begin{figure}[tbp]
 \centering  
 \subfigbottomskip=-2pt 
 \subfigcapskip=-2pt 
 \subfigure[Log Parsing]{
  \includegraphics[width=0.967\linewidth]{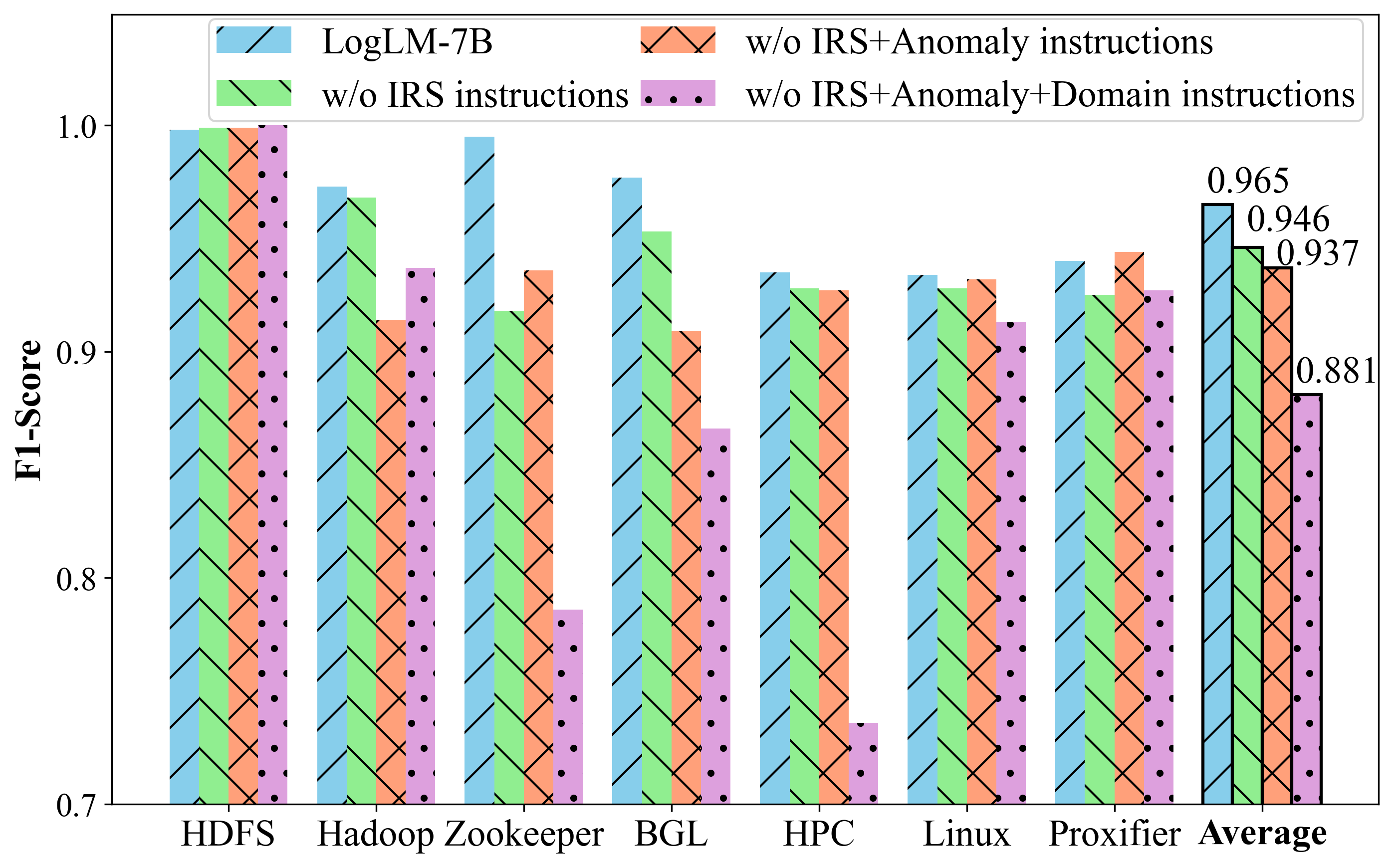}}
   \\
 \subfigure[Anomaly Detection]{
  \includegraphics[width=0.96\linewidth]{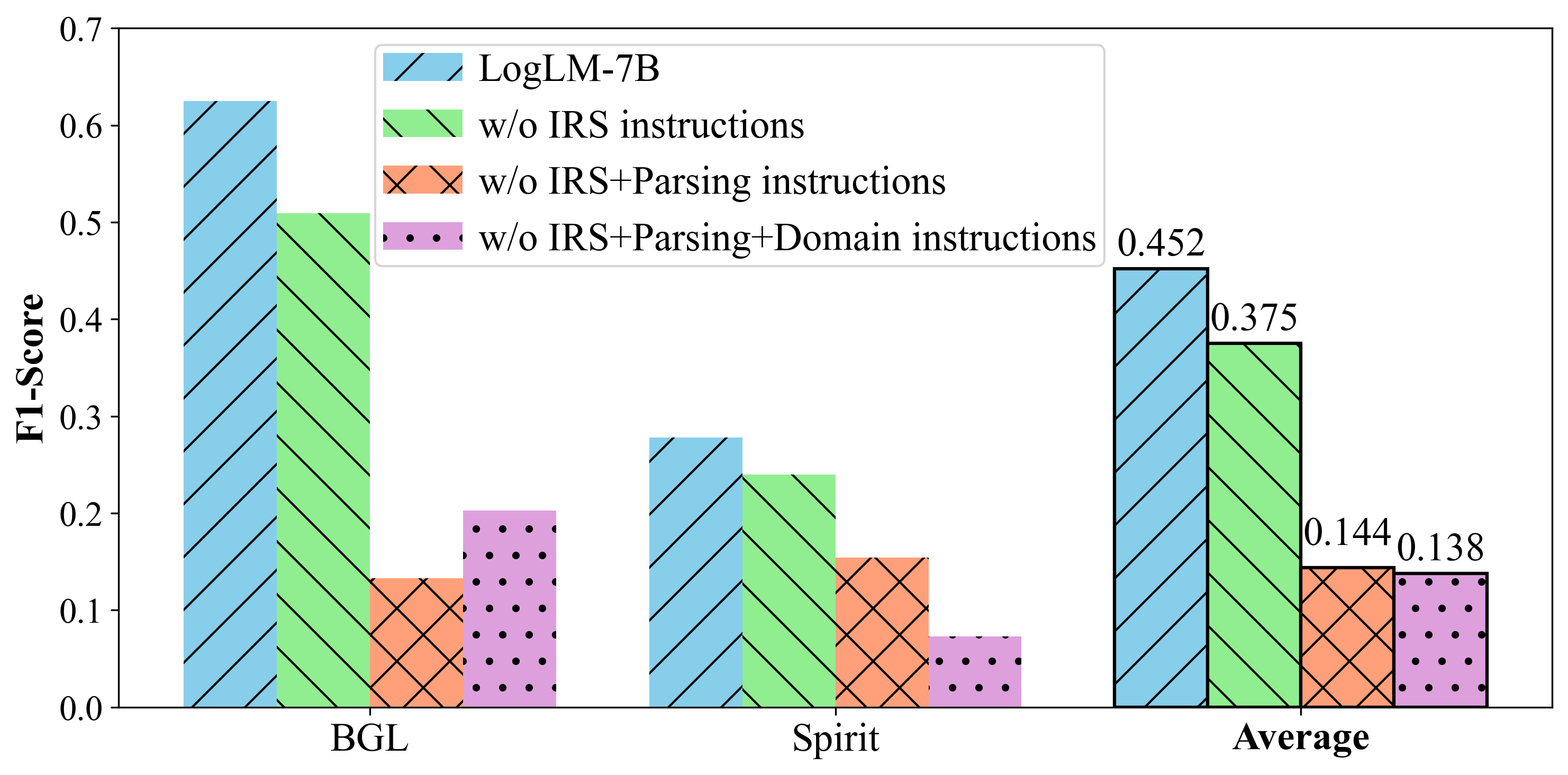}}
 \caption{Ablation study on the training data of LogLM, evaluated on (a) \textit{Log Parsing} and (b) \textit{Anomaly Detection}. \textbf{See additional results (other three tasks, and an upsampling group to control quantity) in our GitHub Page.}}
\label{ablation_exp}
\end{figure}

\paragraph{Result}

As shown in Table~\ref{tab:IRS_exp}, LogLM-7B outperforms existing LLMs in all three capabilities. The advantages in ROUGE-1 and ROUGE-2 indicate that LogLM's responses recalled more key information in the reference responses, which is essential for providing correct root causes of logged errors and feasible solutions. Also, the higher scores in ROUGE-L indicate the sentence structure of LogLM's responses are more similar to human-calibrated reference responses derived from community expert resolutions. This suggests that LogLM's answers are more natural in style and layout, which is especially advantageous for generating human-preferred interpretations of logs. Moreover, a significant advantage of LogLM-7B in BLEU score is observed across the three capabilities, indicating that LogLM generated more precise and readable contents with less hallucination.

\subsection{RQ2: Ablation Study on Training Data Composition}\label{sec:RQ2}

\subsubsection{Evaluation Setting}

In this section, we trained models from LLaMA-2-7B with the same training settings, except using four distinct groups of instruction pairs, and evaluated their performances on \textit{Log Parsing} and \textit{Anomaly Detection}. The four groups are as follows: (1) LogLM: This group utilizes the complete instruction dataset in Table~\ref{tab:traing_set}, resulting in the exact LogLM-7B model. (2) W/o IRS instructions: In this group, instructions related to \textit{Log Interpretation}, \textit{Root Cause Analysis}, and \textit{Solution Recommendation} were excluded from the training process. (3) W/o IRS+Parsing instructions (or w/o IRS+Anomaly instructions): In addition to the IRS-related instructions, this group further excludes instructions involving \textit{Log Parsing} (or \textit{Anomaly Detection}) to assess the model's ability to perform \textit{Anomaly Detection} (or \textit{Log Parsing}). In this case, only task-relevant instruction pairs remain in the training set. (4) W/o IRS+Parsing+Domain instructions (or w/o IRS+Anomaly+Domain instructions): This group involves the further exclusion of instruction pairs containing logs from other domains, resulting in a training set composed solely of in-domain logs relevant to the specific task.

\subsubsection{Result}

As shown in Fig.~\ref{ablation_exp}, the average performance for both tasks consistently declines when instruction pairs from other tasks and domains are excluded. Since the aim of instruction tuning is to train the model to generate human-preferred responses to user instructions, incorporating diverse instruction pairs from various tasks and domains enhances the model's alignment with human preferences. Rather than merely fitting on task labels from a specific domain, the model is trained to leverage its pre-existing knowledge and inference capabilities to address a wide range of log-related problems, thereby developing a robust and comprehensive log analysis capability while reducing dependency on in-domain task data. This suggests LogLM's potential to benefit from the continual emergence of new logs and tasks in real-world systems.

\subsection{RQ3: Generalization on Unseen \& Complex Instructions}\label{sec:RQ3}

\subsubsection{Performing Unseen Analysis Tasks}\label{sec:unseen_task}

\begin{table}[tbp]
\caption{Evaluation of LogLM on Tasks Unseen from Training}
\resizebox{\linewidth}{!}{
\begin{tabular}{@{}l@{\hskip 0.05in}c@{\hskip 0.05in}c@{\hskip 0.05in}c@{\hskip 0.05in}c@{\hskip 0.05in}}
\toprule
Unseen Task$^{\mathrm{a}}$           & BLEU           & ROUGE-1         & ROUGE-2               & ROUGE-L         \\ \midrule
Log & \textbf{6.451} & \textbf{36.841} & \textbf{13.212}       & \textbf{25.364} \\
Interpretation                                   & (v.s. 4.466)$^{\mathrm{b}}$   & (v.s. 30.512)   & (v.s. 11.786)         & (v.s. 19.123)   \\\midrule
Root Cause      & \textbf{9.253}          & \textbf{36.089} & \textbf{14.747}       & \textbf{26.890} \\
Analysis                                    & (v.s. 2.563)   & (v.s. 25.824)   & (v.s. 7.909)          & (v.s. 15.221)   \\\midrule
Solution       & 1.233          & \textbf{24.221} & 6.033                 & \textbf{17.475} \\
Recommendation                                    & (v.s. \textbf{2.698})   & (v.s. 23.618)   & (v.s. \textbf{6.936}) & (v.s. 15.574)   \\ \bottomrule
\multicolumn{5}{l}{$^{\mathrm{a}}$ Unseen task $X$: testing on $X$ after excluding $X$ from training.}\\
\multicolumn{5}{l}{$^{\mathrm{b}}$ Compare LogLM with best baselines of the task from Table~\ref{tab:IRS_exp}.}\\
\end{tabular}
}\label{tab:unseen_task}
\end{table}

Table~\ref{tab:unseen_task} displays the performance of LogLM on unseen tasks, which were excluded from training set in advance (\emph{i.e.}, four capabilities for training and the remaining one capability for testing). A universal decline on performance of corresponding capabilities can be observed when being treated as the unseen task. The capability with the sharpest performance drop is \textit{Solution Recommendation}, possibly due to its challenging nature compared with other capabilities. However, despite not being trained on the excluded task, LogLM still outperforms existing LLMs (highest scores from Table~\ref{tab:IRS_exp} achieved by baselines are displayed in the parentheses) in 10 out of the 12 test terms. This result not only indicates a strong generalization ability of LogLM on unseen instructions, but also suggests that LogLM acquired a comprehensive problem-solving ability related to log analysis through capturing cross-task connections. 

\subsubsection{Following Complex User Instructions}\label{sec:complex_instruction}

Fig.~\ref{fig_user_instruction} shows three cases in which LogLM-7B responds to complex user instructions. In case (a), an open-ended question related to log analysis is posed, seeking advice on improving logging strategies. LogLM’s response includes several recommendations, such as incorporating timestamps, establishing log levels, and adding concise messages, demonstrating that LogLM possesses a solid understanding of log analysis and can effectively apply log-related knowledge to generate appropriate responses. In case (b), LogLM is tasked with a novel problem: comparing two logs and identifying the differences. The model successfully identifies key distinctions, such as the use of proxies and encryption methods, indicating that LogLM's analytical capabilities are not confined to predefined tasks or domains but can extend to emerging problems with logical accuracy. Case (c) involves a combination of two known capabilities from training: \textit{Anomaly Detection} and \textit{Log Interpretation}, requiring the model to detect anomalies while providing justification for its conclusions. This type of instruction is common in real-world scenarios where O\&M engineers need more than just predicted values from a model—they require detailed explanations to inform further actions~\cite{liu2024logprompt}. In response, LogLM effectively follows the instruction, delivering both a conclusion and a well-reasoned justification for the ``10061 error''.

The flexibility exhibited by LogLM may be attributed to the systematic composition of instruction pairs in Table~\ref{tab:traing_set}, which span two log-understanding and three log-application capabilities. Consequently, through training on these diverse pairs of user instructions and human-preferred responses, LogLM has developed not only a deep comprehension of log semantics and structures but also a robust capacity to address practical log-related problems and generate well-formatted responses. This instruction-following ability highlights the model's potential in dynamic and complex industrial environments.

\section{Discussion}
\subsection{LogLM in Practice}
We further examine the feasibility of our approach in industrial applications, by deploying LogLM as the log analysis component within a software and network O\&M platform of Huawei. The deployment process involved instruction tuning based on Huawei's proprietary LLM, Pangu~\cite{ren2023pangu}, using the proposed instruction dataset in Table~\ref{tab:traing_set}. To support retrieving real-time logs from local devices via private API, instruction pairs involving function calling~\cite{schick2024toolformer} were also added to the instruction dataset as a new capability of LogLM. The fine-tuned model is denoted by LogLM-Huawei.

In practice, the platform operates by allowing users to issue instructions, which are then responded by LogLM-Huawei. The model first retrieves relevant logs from the specified devices by calling system APIs. Upon receiving the logs, LogLM-Huawei performs a thorough analysis and responds to users. The instructions from users span a variety of operational needs, such as resource checks and fault diagnosis.

For instance, a typical query might involve a client asking for an investigation into whether a specific user experienced issues over the past three days. LogLM-Huawei firstly requests an API calling to retrieve logs within past three days from the user's associated devices. The platform will parse the request, collect the logs retrieved from the API and send them to the model. LogLM-Huawei then evaluates the devices' statuses through examining on multiple logs, and provides analysis results on possible issues the user experienced, along with recommendations on actions. Other frequent user instructions include inquiries about specific device statuses and system-generated logs, and related knowledge-based questions.

Since its deployment in the platform managing arrays of devices, LogLM has processed over 30k queries over six months, averaging 200+ queries per day. LogLM's robust log analysis and adaptive instruction-following abilities make it the central intelligence of the O\&M platform, driving real-world applications and delivering valuable insights to customers.

\subsection{Threats to Validity}

Our study has several limitations:

(1) Hallucination in LLMs: Despite the competitive performance of LogLM-7B in log analysis benchmarks, occasional hallucinations observed in generated outputs may affect accuracy. While hallucination is a known issue in LLMs and has garnered attention from AI researchers~\cite{zhang2023siren}, mitigation strategies exist for industrial deployment, such as incorporating post-processing modules, optimizing prompts, or utilizing larger foundation models.

(2) Fairness of Experimental Comparisons: In Section~\ref{sec:RQ1}, the training data for LogLM differs from that of baseline methods, raising concerns about the fairness of comparisons. However, due to the proprietary and pre-trained nature of many baselines (\emph{e.g.}, relying on ChatGPT), identical data usage is unfeasible. To ensure fairness, as discussed in Sections~\ref{sec:logParsing_exp_setting} and \ref{sec:anomaly_exp_setting}, LogLM uses the same or fewer in-domain and task-specific log-label pairs than existing methods for \textit{Log Parsing} and \textit{Anomaly Detection}. Furthermore, for the three IRS capabilities, Table~\ref{tab:unseen_task} shows that LogLM outperforms most baselines, even without task-specific training data.

(3) Biases in Studied Logs:  Most studied logs in this paper are system logs, from middlewares, frameworks to operation systems. In practice, logs from software applications are also analyzed, which may leads to biases in LogLM. However, the availability of open-source datasets for application logs is limited. To address this, logs from the Apache server application are included in the study, mitigating this issue.

(4) Limited Practice Verification: In practice, LogLM has been verified only on Huawei's O\&M platform. Other organizations have different environments and requirements, which may introduce potential external threats to LogLM. Expanding its deployment across various organizations would enhance its generalizability and help mitigate external threats

\section{Conclusion}

In this paper, we present LogLM, an instruction-based log analysis approach. Through our empirical study, three key advantages of LogLM over existing task-based methods are highlighted: (1) greater flexibility, (2) multi-task and generalization capability and (3) enabling single-model deployment. The released instruction dataset of LogLM can facilitate future endeavors in building LLMs for software O\&M. Future work include adding more instruction sets, testing with larger foundation models and further deploying LogLM across various organizations to validate its generalizability.

\clearpage

\bibliographystyle{./IEEEtran}
\bibliography{./mybib}

\end{document}